\documentclass[12pt]{article}
\usepackage{authblk}
\usepackage[bookmarksnumbered, colorlinks, plainpages]{hyperref}
\usepackage{amsmath, amsthm, amscd, amsfonts, amssymb, graphicx, color, booktabs}
\usepackage{xspace}
\numberwithin{equation}{section}
\definecolor{email}{rgb}{0.00,0.00,0.84}
\begin{document}
\setcounter{page}{1}

\title{\large \bf 12th Workshop on the CKM Unitarity Triangle\\ Santiago de Compostela, 18-22 September 2023 \\ \vspace{0.3cm}
\LARGE Measurement of the CP violating phase $\phi_s$ and $\phi_s^{sq\bar{q}}$ at LHCb }

\author[1]{Melissa Maria Cruz Torres  on behalf of the LHCb collaboration \\ email: \href{mailto:melissa.cruz@unah.edu.hn}{melissa.cruz@unah.edu.hn}} 
\affil[1]{Universidad Nacional Aut\'onoma de Honduras, Honduras }
\maketitle

\begin{abstract}
 Precise measurements of the $B_s^0 - \overline{B}_s^0$ mixing parameters provide a powerful test of the Standard Model, offering potential hints to new physics. The LHCb collaboration has performed updated measurements of the $CP$-violating phases $\phi_s^{c\bar{c}s}$ and $\phi_s^{s\bar{s}s}$, which supersede previous results. Also, an alternative approach to determine $\Delta \Gamma_s$ is presented, bringing a new tool that may help to resolve the tension observed between measurements made in $B_s^0 \to J/\psi \phi$ by LHC experiments.

\end{abstract} \maketitle

\section{Introduction}
\noindent One of the key goals of the LHCb experiment is the  measurement of the $CP$-violating phase, $\phi_s$, that originates from the interference of the mixing and direct decay of $B_s$ mesons to $CP$ eigenstates. Within the Standard Model $\phi_s$ is predicted to be equal to $-2\beta_s$, where $\beta_s \equiv arg[-(V_{ts}V_{tb}^*)/(V_{cs}V_{cb}^*)]$, ignoring subleading penguin contributions, and where $V_{ij}$ represents the Cabibbo-Kobayashi-Maskawa matrix elements~\cite{CKM}. In this sense, precise measurements of this phase enhance the potentiality to probe physics beyond the Standard Model \cite{Buras}

\section{Measurement of $\phi_s$  in $B_s^0 \to J/\psi K^+K^- $ decays }

The study of the time-dependent $CP$ asymmetry in decays modes with transitions of the form $b\to c\bar{c}s$ has been performed by several experiments. It provides insights into understanding the $CP$ violation phenomena. The world average value of the $CP$-violating phase, $\phi_s^{c\bar{c}s}$, is found to be $-0.049 \pm 0.019$ rad~\cite{PDG}, which is dominated by the LHCb results in the decay channel $B_s^0 \to J/\psi h^+h^- $ ($h = K$ or $\pi$)~\cite{phis}. 
An update of the phase, $\phi_s^{c\bar{c}s}$, as well of the physics parameters $|\lambda|$, $\Delta \Gamma_s$, $\Delta \Gamma_s - \Gamma_d$ and the $B_s^0$ mass difference $\Delta m_s$, have been performed using the golden channel $B_s^0 \to J/\psi K^+K^- $  in the vicinity of $\phi(1020)$, which is reported in this talk. These results supersede the previous ones. The dataset used includes the full data sample from 2015 to 2018, collected by the LHCb detector corresponding to an integrated luminosity of 6 fb$^{-1}$ at $\sqrt{s}$ = 13 TeV.

The analysis strategy consists in performing a flavour-tagger time-dependent angular analysis, where four polarization states are identified, namely $A_0, A_{||}, A_{\perp}$ for the P-wave  and $A_s$ for the S-wave, regarding to the polarization states of the $K^+K^-$ system. Candidates with $K^+K^-$ invariant masses in the range [900,1050] MeV/$c^2$  from $B_s^0 \to J/\psi(\to \mu^+\mu^-)K^+K^-$ decays are selected following the strategy described in reference \cite{phis}. Experimental improvements benefit the updated measurements, as to be the flavour-tagging algorithms, the $B_s^0$ decay-time resolution model as well as in the determination of the particle identification. 
A gradient-boosted decision tree (BDT) classifier is trained separately for each year from 2016 to 2018, improving the signal to background ratio by a factor of 50. The Cross-feed contamination due to pion and proton misidentification is handled by appropiate particle identification requirements and mass constraints. 

\begin{figure} [hbt!]
\centering
\includegraphics[width=0.9\textwidth]{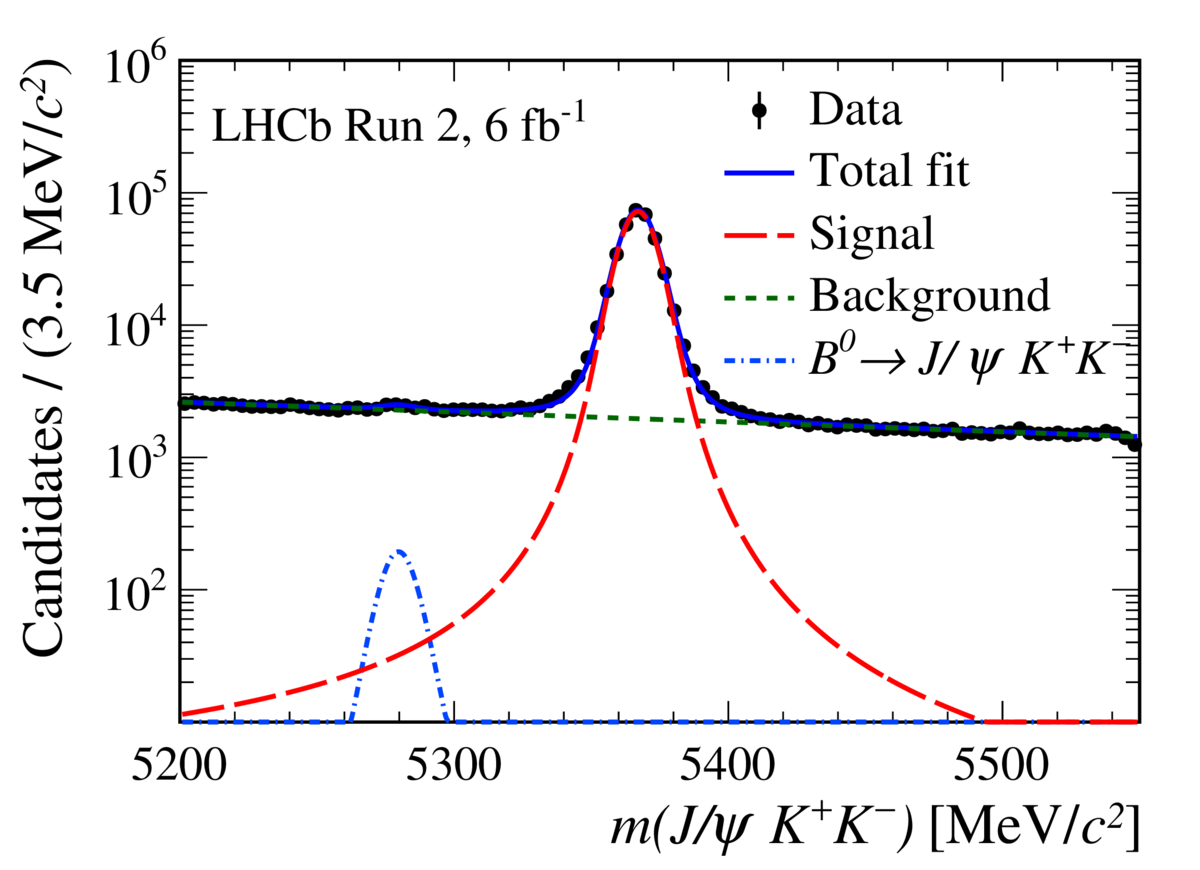 }
 \caption{Invariant mass distribution of $B_s^0 \to J/\psi K^+K^- $ for the full data sample. The projection of the maximum likelihood fit is overlaid~\cite{phis2024}.} 
\label{fig:mass}
\end{figure}

The data sample is divided into 48 independent sub samples, corresponding to six bins in the $\phi(1020)$ region, two trigger configurations and four year of data taking. By performing a maximum likelihood fit, a total yield of about 349 000 signal decays are extracted. The fit to the invariant mass is shown in Fig.~\ref{fig:mass}.

\begin{figure} [hbt!]
\centering
\includegraphics[width=0.45\textwidth]{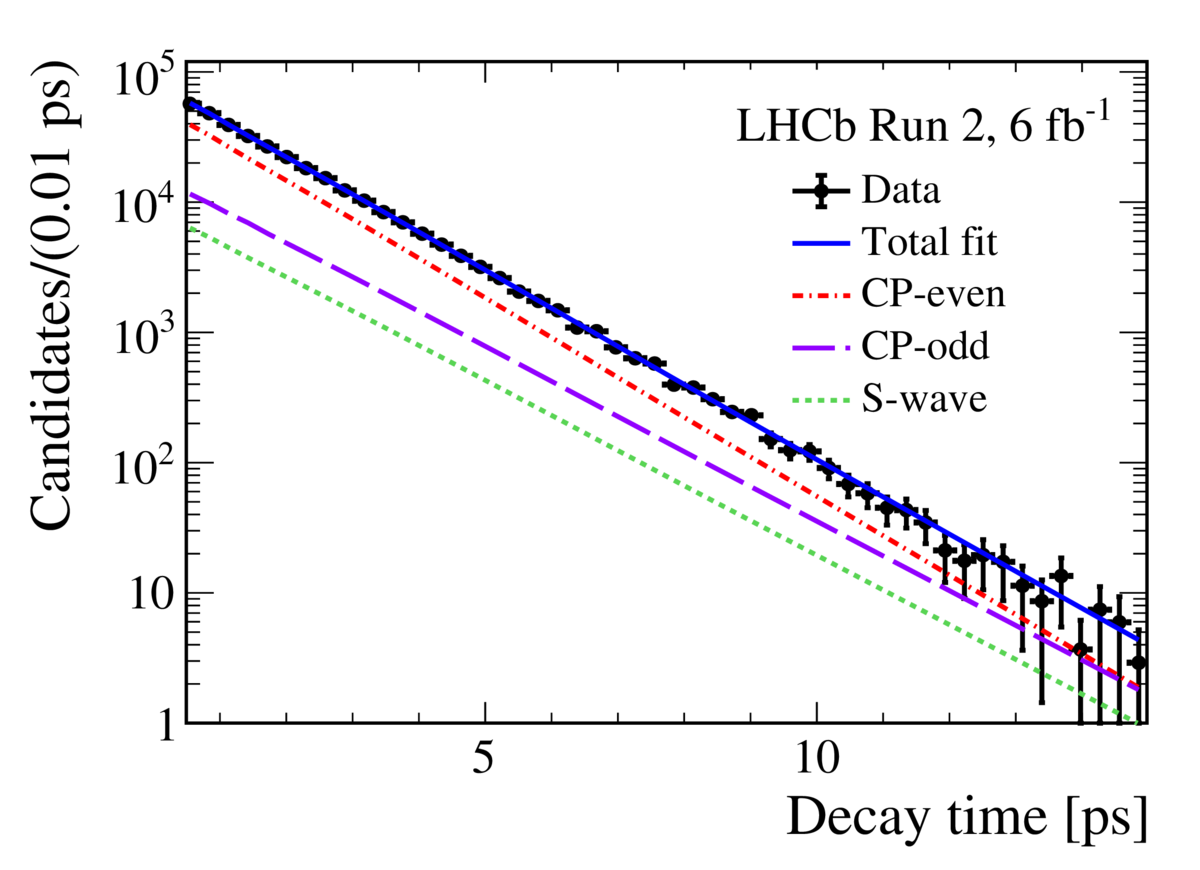}
\includegraphics[width=0.45\textwidth]{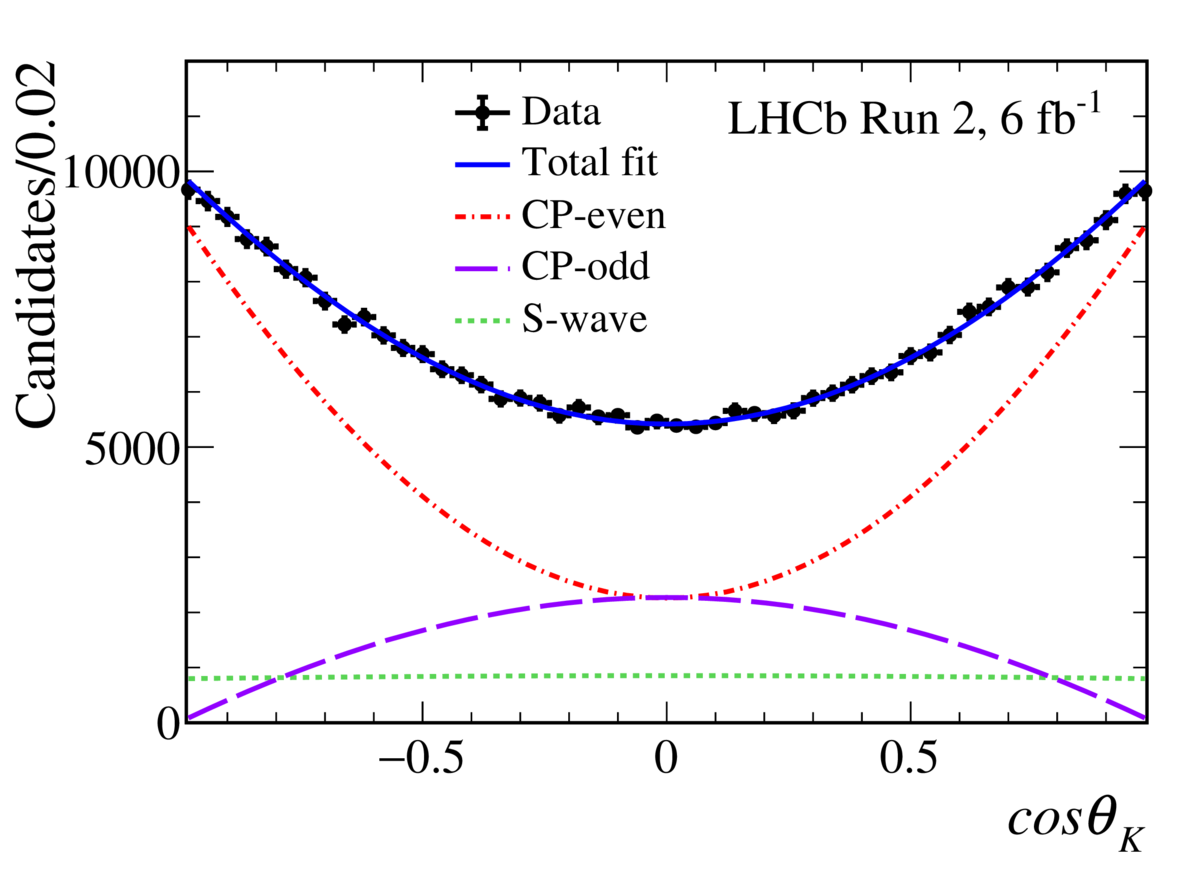}
\includegraphics[width=0.45\textwidth]{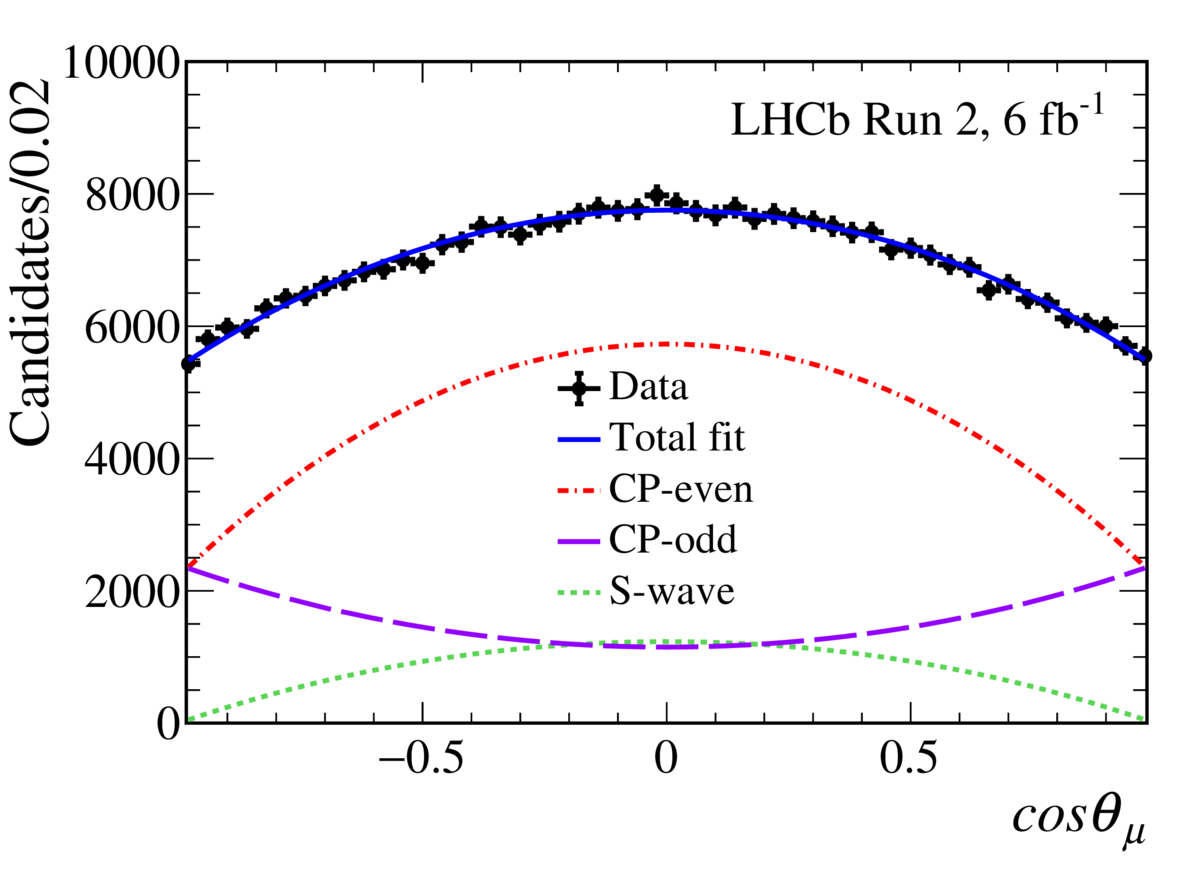}
\includegraphics[width=0.45\textwidth]{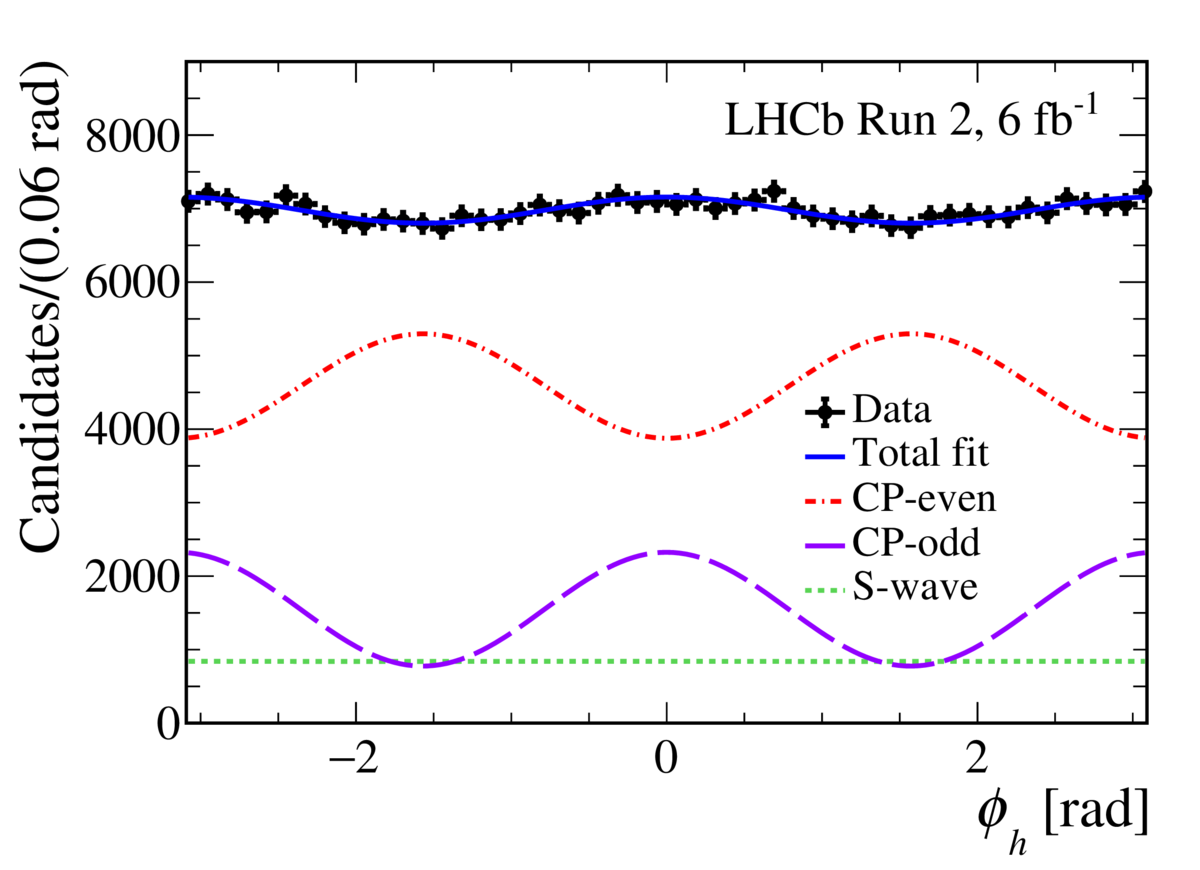}

 \caption{Background-substracted data distribution with fit projections for decay-time and decay-angles~\cite{phis2024}.} 
\label{fig:Fig1}
\end{figure}

To determine the phase $\phi_s$ in $B_s^0 \to J/\psi K^+K^- $, a weighted simultaneous fit to the distributions of decay time and decay angles (cos$\theta_K$, cos$\theta_\mu$, $\phi_h$) in the helicity basis is performed in the 48 sub samples. Along with the determination of $\phi_s$, the physics parameters and the polarization amplitudes $A_k = |A_k|e^{-i\delta_k}$ are also determined. The sub-index, $k$, in the latter, stands for polarization states of the $K^+K^-$ system. 
The probability density function for the signal takes into consideration the decay-time and angular efficiencies, decay-time resolution and flavour tagging.
The results from the fit are shown in Table~\ref{tab:Tab1}~\cite{phis2024}. The background subtracted data distribution with fit projections overlaid can be seen in Fig~\ref{fig:Fig1}.

 \begin{table}
 \centering
 \caption{Physics parameters results. The first uncertainty is statistical and the second systematic~\cite{phis2024}}
 \label{tab:Tab1}
    \begin{tabular}{lc}
    \hline
        Parameter  & Result   \\
        \hline
      $\phi_s$ [rad] & $-$0.039 \,$\pm$0.022 \,\,$\pm$0.006  \\  
      $|\lambda|$ & \phantom{+}1.001 \,$\pm$0.011 \,\,$\pm$0.005 \\
      $\Gamma_s - \Gamma_d$ [ps$^{-1}$] & \,\, $-$0.0056\,\,$^{\:+0.0013}_{\:-0.0015}$ \,$\pm$0.0014 \\
      $\Delta \Gamma_s$ [ps$^{-1}$] & \phantom{+}\,\, 0.0845$\pm$0.0044\,$\pm$0.0024\\
      $\Delta m_s$ [ps$^{-1}$] & \,\,\, 17.743\,$\pm$0.033 \,\,$\pm$0.009 \\
      $|A_{\perp}|^2$ & \phantom{+}\,\, 0.2463$\pm$0.0023\,\,$\pm$0.0024 \\
      $|A_{0}|^2$ & \phantom{+} \, 0.5179$\pm$0.0017$\pm$0.0032 \\
      $\delta_{\perp} - \delta_0$ [rad] &\phantom{+}2.903\phantom{+}$^{\:+0.075}_{\:-0.074}$  \:\,\,\,$\pm$0.048 \\
      $\delta_{||} - \delta_0$ [rad] & \phantom{+} 3.146  \,$\pm$0.061 \,$\pm$0.052 \\
      \hline
      
    \end{tabular}
 
 \end{table}

\begin{figure} [hbt!]
\centering
\includegraphics[width=0.9\textwidth]{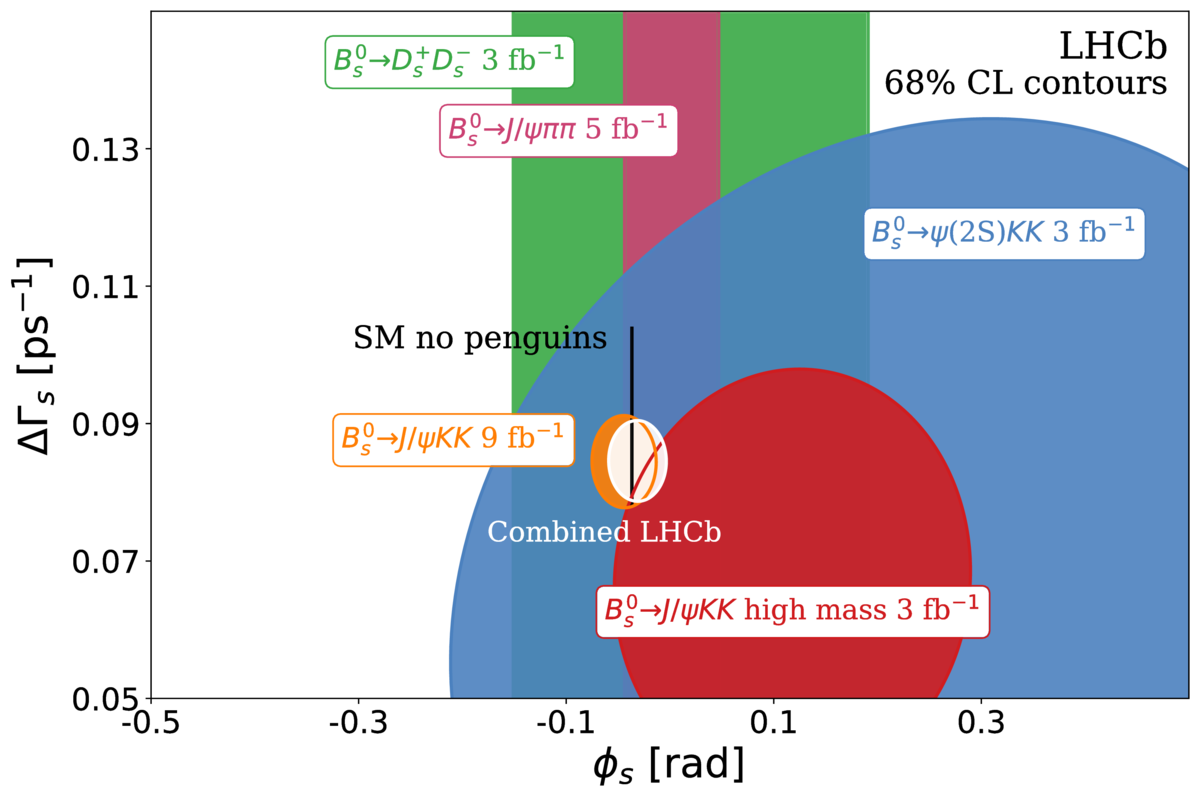 }

 \caption{$\phi_s$ combination for all LHCb results~\cite{phis}.} 
\label{fig:Fig2}
\end{figure}

The results are in good agreement with LHCb Run 1~\cite{phis1} and 2015+2016~\cite{phis} measurements. The obtained values for $\phi_s$, $\Delta \Gamma_s$ and $\Gamma_s -\Gamma_d$ represent the most precise measurements to date and are in good agreement with SM expectations.
No evidence for $CP$ violation in found. Results also show no evidence for polarization dependence of $\phi_s$. Combination of all LHCb $\phi_s$ measurements of $B_s^0$ decays via $b \to c\bar{c}s$ is shown in Figure~\ref{fig:Fig2}. The combined value is $\phi_s = -0.031 \pm 0.018$ rad.

 \section{$CP$ violation measurements in the penguin-mediated decay $B_s^0 \to \phi\phi$}

$B$ mesons decays with Flavour-Changing Neutral Current (FCNC) transitions provides a very sensitive ground to search for new physics. One of the benchmark channels to study at LHCb, in FCNC decays, is the penguin-dominated process $B_s^0 \to \phi\phi$ where $\phi \to K^+K^-$ as loop contributions could reveal new sources of $CP$ violation.
A time-dependent angular analysis, using the full Run 2 dataset, is performed on this channels to determine the $CP$-violating parameters $\phi_s^{s\bar{s}s}$ and $|\lambda|$. Any deviation of their expected values, 0 or 1 respectively, could be an indication of new physics entering in the penguin decay or the $B_s^0$ mixing. The analysis is performed using the full Run 2 data sample.

\begin{figure} [hbt!]
\centering
\includegraphics[width=0.9\textwidth]{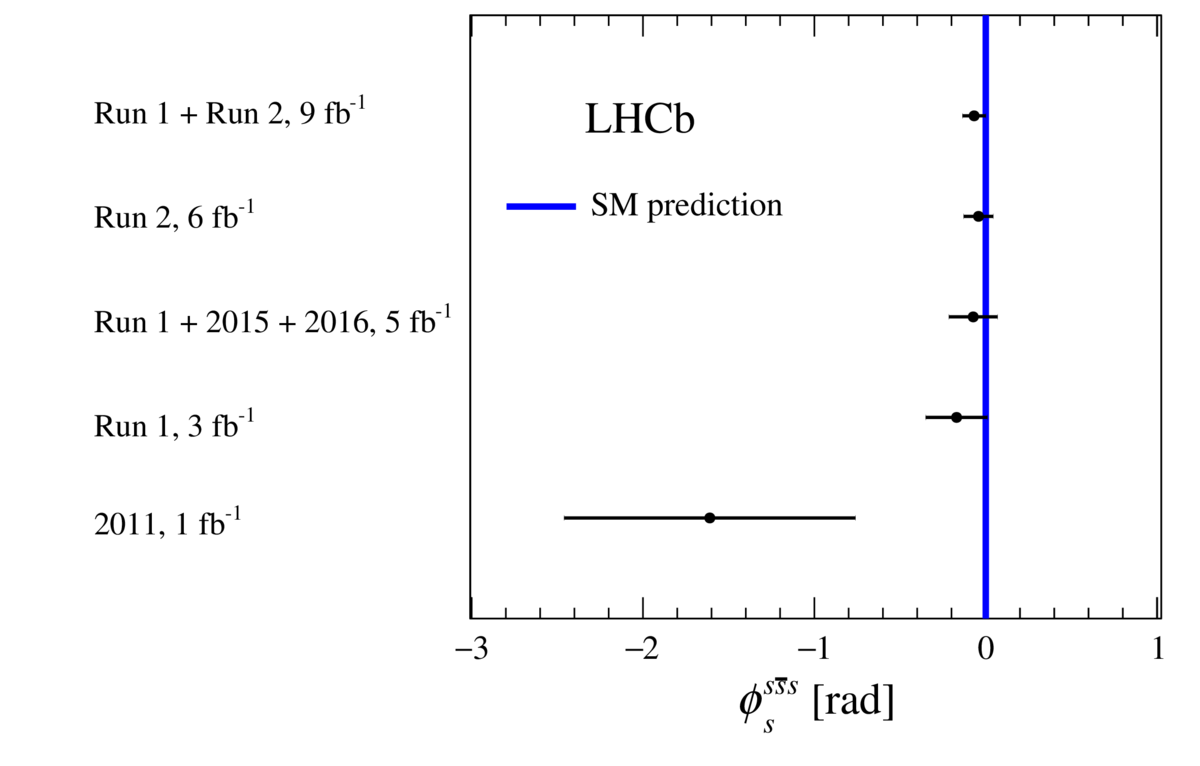}

 \caption{Comparison of $\phi_s^{s\bar{s}s}$ measurements by LHCb experiment. Results presented in this work are the denoted as Run 1 + Run 2, 9 fb$^{-1}$ and Run 2, 6 fb$^{-1}$~\cite{phisss}. The blue vertical line represent the SM prediction.} 
\label{fig:Fig3}
\end{figure}

The $B_s^0 \to \phi\phi$ candidates are selected in the [5150,5600] MeV/$c^2$ mass range, yielding about 16,000 of signal events. The three polarization states $B_s^0$ are considered, namely $A_0, A_{||}$ and $A_{\perp}$ and both, polarization-dependent and polarization-independent, scenarios are evaluated.
The parameters $\phi_{s,i}$ and $|\lambda_{i}|$ are defined by the equation 

\begin{equation}
    \frac{q}{p} \cdot \frac{\overline{A}_i}{A_i} = \eta_{i} |\lambda_{i}| e^{-i\phi_{s,i}},
\end{equation}
where $\eta_i$ is the $CP$ eigenvalue of the polarization state $i$, $q$ and $p$ are complex numbers relating the $B_s^0$ mass eigenstates to the flavour eigenstates. The differential decay rate is written as 
\begin{equation}
    \frac{d^4\Gamma(t, \vec{\Omega})}{dtd\vec{\Omega}} \varpropto \sum_{k=1}^{6} h_k(t) f_s(\vec\Omega),
\end{equation}
where $t$ is the decay time and $\vec \Omega = (\theta_1, \theta_2, \chi$) refer to the helicity angles of the $K^+$ mesons in the $\phi$ rest frame. $\chi$ is the angle between the $\phi \to KK$ decay planes. The angular function $f_k(\vec \Omega)$ and the time-dependent function $h_k(t)$ are defined as given in references \cite{fk} and \cite{hk}, respectively. 
After applying a maximum-likelihood fit to the distribution of $t$, $\vec\Omega$ and the initial $B_s^0$ state, the polarization-independent measurements of the $CP$-violation parameters are reported to be $\phi_s^{s\bar{s}s} = -0.074 \pm 0.069$ rad and $|\lambda| = 1.009 \pm 0.030$ in combination with Run 1 results~\cite{phis}. This is the most precise measurement of $CP$ violation in $B_s^0 \to \phi\phi$ decays to date~\cite{phisss} as can be seen in Fig~\ref{fig:Fig3}. The results solely due the Run 2 data are reported to be $\phi_s^{s\bar{s}s} = -0.042 \pm 0.075 \pm 0.009$ rad and $|\lambda| = 1.004 \pm 0.030 \pm 0.009$, where the first uncertainty is statistical and the second systematic~\cite{phisss}.  The total measurable observables can be seen in Table~\ref{tab:Tab2}.  For the polarization-dependent scenario, no dependence of the $CP$-violation parameters in the polarization states is found.

\begin{table}[hbt]
\centering
\caption{Measured observables in the polarization-independent fit for the full Run 2 dataset~\cite{phisss}.}
\label{tab:Tab2}
\begin{tabular}{lccc}
   \hline 
Parameter&  Result \\
\hline
	   $\phi_{ s\xspace\xspace}^{ s\xspace\xspace\overline  s\xspace\xspace\xspace s\xspace\xspace}\xspace$ [ $\text{\,rad}$ \xspace]              &  $-0.042\pm 0.075 \pm0.009 $\\
	   $|\lambda|$  &                  
          $ \;\;\;1.004\pm0.030\pm0.009$\\
	   $\left| A_{0}\xspace\right|^2$   &            
         $\;\;\;0.384 \pm0.007\pm0.003$\\
         $\left| A_{\perp}\xspace\right|^2$ &              
          $\;\;\;0.310\pm0.006 \pm0.003$\\
          $\delta_{\parallel}\xspace-\delta_{0}\xspace$ \;[ $\text{\,rad}$ \xspace]  &  
         $ \;\;\;2.463\pm0.029 \pm0.009$\\
	   $\delta_{\perp}\xspace-\delta_{0}\xspace$ \;[ $\text{\,rad}$ \xspace]&    
          $ \;\;\;2.769\pm0.105\pm0.011$\\  
   \hline
\end{tabular}
\end{table}

\section{A measurement of $\Delta \Gamma_s$ from $B_s^0 \to J/\psi \eta^{'}$ and $B_s^0 \to J/\psi \pi^+\pi^-$ decays}

As explained before, the measurement of the $B_s^0 - \overline{B}_s^0$ mixing parameters offer a powerful test of the Standard Model. Particularly, $\Delta \Gamma_s$ have been determined experimentally using the golden channel $B_s^0 \to J/\psi \phi$ by ATLAS~\cite{ATLAS}, CMS~\cite{CMS} and LHCb~\cite{phis2024} experiments. The results are precise but in tension with each other. In this analysis, an alternative approach to determine $\Delta \Gamma_s$ that follows closely the formalism proposed in~\cite{Fleischer},  is presented.

Given that $CP$-odd modes measure the heavy mass eigenstate lifetime ($\tau_H = 1/\Gamma_H$) and $CP$-even modes measure  the light mass eigenstate lifetime  ($\tau_L = 1/\Gamma_L$), $\Delta \Gamma_s$ can be determined from the decay-width difference between a $CP$-odd and a $CP$-even $B_s^0$ state.
In this study it is used the $CP$-even decay $B_s^0 \to J/\psi \eta^{'}$ and the $CP$-odd decay $B_s^0 \to J/\psi \pi^+\pi^-$, where $J/\psi \to \mu^+\mu^-$ and $\eta^{'} \to \rho^0\gamma$.

If $CP$ violation is considered negligible the time dependent rate can be expressed as 
\begin{equation}
    \Gamma(B_s^0(t) \to f) \varpropto e^{-\Gamma_s t} [\rm{cosh}(\frac{\Delta \Gamma_s t}{2}) +\eta_{CP} \rm{sinh}(\frac{\Delta \Gamma_s t}{2}) ],
    \label{eq:Eq1}
\end{equation}
where $\eta_{CP}$ is $1$ or $-1$ for $CP$-odd or $CP$-even states, respectively. By integrating Eq.~\ref{eq:Eq1} over a time range $[t_1,t_2]$, and performing their ratio, it is obtained the following equation

\begin{equation}
    R_i = \frac{N_L}{N_H} \varpropto \frac{[e^{-\Gamma_s t(1+y)}]_{t_1}^{t_2}}{[e^{-\Gamma_s t(1-y)}]_{t_1}^{t_2}} \cdot \frac{(1-y)}{(1+y)},
    \label{eq:Eq2}
\end{equation}
the parameter $y$ is defined to be $\Delta \Gamma_s/2\Gamma_s$ and experimental corrections are taken into account into; $R_i$ is corrected by the relative efficiency in each decay time bin as $R_i = A_i \frac{N_L^{RAW}}{N_H^{RAW}}$. A total of 8 time bins are chosen, whose ranges are determined using simulation samples. Similar yields are expected in each bin and it follows the purpose of optimizing the sensitivity of $\Delta \Gamma_s$. Finally, $\Delta \Gamma_s$ is determined from $\chi^{2}$ minimization of the expression above (Eq.\ref{eq:Eq2}), where $\Delta \Gamma_s$ and a normalization factor are free parameters.

\begin{figure}[hbt!]
\centering
\includegraphics[width=\textwidth]{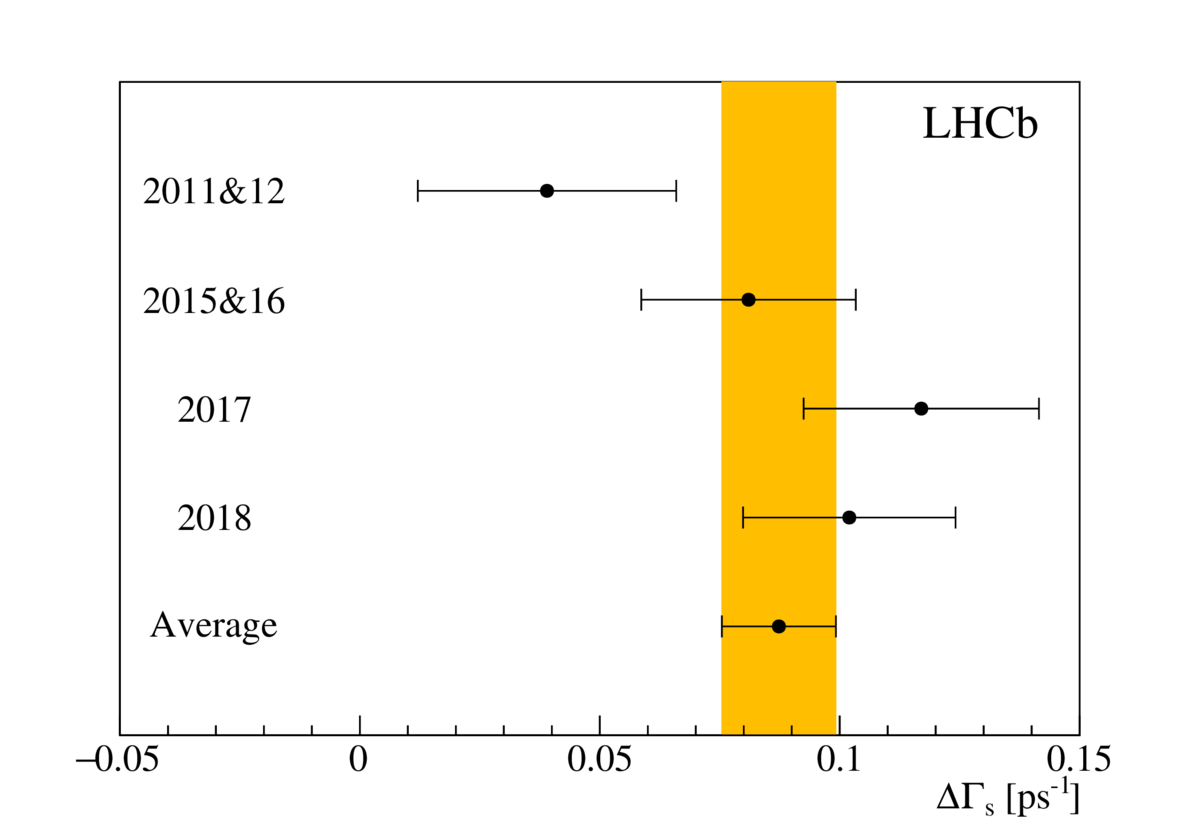}

 \caption{Measurements of $\Delta \Gamma_s$ for the sets of data taking and the weighted average. The orange band represents $1\sigma$ error~\cite{Gammas}}. 
\label{fig:Fig4}
\end{figure}

The results are presented in Table~\ref{tab:Tab3}. The weighted average is reported to be $\Delta\Gamma_s = 0.087 \pm 0.012 \pm 0.0009$ ps$^{-1}$, where the first uncertainty is statistical and the second systematic~\cite{Gammas}. The study was performed using the full data sample of proton-proton collisions collected by the LHCb between 2011 and 2018 corresponding to a integrated luminosity of 9 fb$^{-1}$ in the center-of-mass energy $\sqrt{s}$ = 7, 8 and 13 TeV. The comparison of the results by years of datasets are shown in Fig.~\ref{fig:Fig4}.

In summary, the value obtained for $\Delta \Gamma_s$ is in agreement with the HFLAV average value, $\Delta \Gamma_s = 0.074 \pm 0.006$ ps$^{-1}$~\cite{hflav1}, obtained from time-dependent angular analyses of $B_s^0 \to J/\psi \phi$, where the initial flavour of the state is tagged. The value reported also agrees with the HFLAV average, $\Delta \Gamma_s = 0.083 \pm 0.005$ ps$^{-1}$~\cite{hflav1}, that includes constraints from other untagged effective lifetime measurements. $\Delta \Gamma_s$ is measured for the first time using using the decay mode $B_S^0 \to J/\psi \eta^{'}$ .

 \begin{table}[hbt]
 \centering
 \caption{$\Delta \Gamma_s$ results and $\chi^2$ probability for the four datasets~\cite{Gammas}.}
 \label{tab:Tab3}
    \begin{tabular}{lcc}
        Dataset  & $\Delta \Gamma_s$ [$\text{\,ps}\xspace^{-1}$] & P($\chi^2$) \\
        \hline
      2011$\&$12 & 0.039 $\pm$ 0.026 & 0.83 \\
      2015$\&$16 & 0.081 $\pm$ 0.022 & 0.77 \\
      2017 & 0.117 $\pm$ 0.024  & 0.57 \\
      2018 & 0.102 $\pm$ 0.021 & 0.78 \\
    \end{tabular}
 
 \end{table}

\bibliographystyle{amsplain}

\end{document}